\documentclass[preprint2]{aastex6}

\begin{document}

\title{On Lithium-Rich Red Giants. II. \\
 Engulfment on the Giant Branch of Trumpler 20}


\author{
Claudia Aguilera-G\'omez\altaffilmark{1,2}, 
Julio Chanam\'e\altaffilmark{1,2},
Marc H. Pinsonneault\altaffilmark{3,4}, and
Joleen K. Carlberg\altaffilmark{5,6}
}
\altaffiltext{1}{Instituto de Astrof\'isica, Pontificia Universidad Cat\'olica de Chile,  Av. Vicu\~na Mackenna 4860, 782-0436 Macul, Santiago, Chile.}
\altaffiltext{2}{Millennium Institute of Astrophysics, Santiago, Chile.}
\altaffiltext{3}{Department of Astronomy, The Ohio State University, Columbus, OH 43210, USA.}
\altaffiltext{4}{Center for Cosmology and Astroparticle Physics, The Ohio State University, Columbus, OH 43210, USA.}
\altaffiltext{5}{NASA Goddard Space Flight Center, Code 667, Greenbelt MD 20771 USA.}
\altaffiltext{6}{NASA Postdoctoral Program Fellow.}

\begin{abstract}
 
The $Gaia$-ESO survey recently reported on a large sample of lithium (Li) abundance determinations for evolved stars in the rich open cluster Trumpler 20. They argue for a scenario where virtually all stars experience post main sequence mixing and Li is preserved in only two objects. We present an alternate explanation, where Li is normal in the vast majority of cluster stars and anomalously high in these two cases.  We demonstrate that the Li upper limits in the red giants can be explained with a combination of main sequence depletion and standard dredge-up, and that they are close to the detected levels in other systems of similar age. In our framework, two of the detected giants are anomalously Li-rich, and we propose that both could have been produced by the engulfment of a substellar mass companion of $16^{+6}_{-10}\,\mathrm{M_J}$. This would imply that $\sim5\%$ of $1.8\,\mathrm{M_\odot}$ stars in this system, and by extension elsewhere, should have substellar mass companions of high mass that could be engulfed at some point in their lifetimes. We discuss future tests that could confirm or refute this scenario.

\end{abstract}

\keywords{stars: abundances --- 
stars: chemically peculiar --- planet-star interactions --- open clusters and associations: individual: Trumpler 20}

\section{Introduction} 

When stars enter their red giant branch (RGB) evolution, they go through some important changes: the energy production changes from the core to a shell surrounding it and the envelope of the stars expands. Also, the surface composition changes due to the first dredge-up (FDU), the deepening of the convective layer that brings nuclear-processed material from the stellar interior to the surface. If the element is depleted in the stellar interior, its surface abundance will be diluted during the FDU. This is the case of lithium (Li). If stars enter the RGB with a solar system meteoritic abundance of $\mathrm{A(Li)}$\footnote{$\mathrm{A(Li)}=log(N_{Li}/N_{H})+12.00$, where $N_x$ is the number of atoms of element ``$x$"}$=3.3$, the Li abundance post-FDU should be $\mathrm{A(Li)}<1.5$. After FDU, no further abundance changes are expected according to standard models.

However, it is well documented observationally that there is an extra-mixing process that changes the surface abundance of giants triggered after the RGB bump \citep[e.g.,][]{gratton00, lind09}.

Also, although they are uncommon, Li-rich giants do exist \citep[e.g.,][]{wallersteinsneden82,brown89}. Some explanations have been proposed for these atypical Li-rich giants. One of these is internal production through the Cameron-Fowler chain \citep{cameronfowler71}, which additionally requires extra-mixing mechanisms to transport the freshly produced Li into cooler regions of the star, preventing it from burning by proton capture. Other possibilities are external to the star. These include the accretion by the star of a planet or brown dwarf \citep{alexander67}, which have not burned Li during their lifetimes, or mass transfer from an evolved AGB star, where Li can be produced by hot bottom burning under convective conditions and thus get transported to the outer layers \citep{sackmannboothroyd92}.

A key piece of information for distinguishing among mechanisms of Li enhancement is the evolutionary phase of the enriched giants. Recent observations indicate that Li-rich giants are found all along the RGB \citep[e.g.,][]{lebzelter12}, as well as the clump \citep{kumar11,SA14, monaco14, reddylambert16}, so there seems that Li-rich giants are not restricted to a particular evolutionary phase. Unfortunately, the exact mass and evolutionary stage of giants can be tricky to obtain when they are located in the field and have no parallax or asteroseismic data, as it is the case for most of the known Li-rich giants. 

This difficulty can be partially overcome in clusters, where the giants have similar masses, and, since the cluster members share the same original composition, distance, and age, differences in giants' surface abundances are tied to internal processes acting inside stars. Open clusters have an additional advantage over globular clusters, as they do not present evidence of multiple populations so far.

In a recent work, \citet{smiljanic16} (S16 hereafter) studied the Li abundance of giants in the open cluster Trumpler 20, in the context of the $Gaia$-ESO Survey \citep{randichgilmore13}. S16 find two giants with a high Li abundance when compared to the rest of the cluster. Thus, this sample offers a great opportunity to study extra-mixing and the Li-enrichment phenomenon.

This letter is organized as follows. In Section 2 we re-analyze the Trumpler 20 data published by S16, and critically discuss their interpretation and its implications. We offer an alternative explanation based on our earlier work on planet/brown dwarf engulfment \citep{AG16} in Section 3, concluding in Section 4.

\section{The possibility of extra mixing in the Trumpler 20 giants}\label{s_mixing}

The $Gaia$-ESO survey presents observations for 41 stars in the open cluster Trumpler 20, with an age of $\sim 1.6$ Gyrs and [Fe/H]=$+0.17$, that are likely members based on radial velocities \citep{donati14}. Carbon and nitrogen abundances for those stars were obtained by \citet{tautvaisiene15}, and the Li abundance including corrections by nonlocal thermodynamical equilibrium, as well as the atmospheric parameters we use throughout this letter, are those reported by S16.

S16 measured Li in 40 giants and 1 subgiant. Only 4 of the 40 giants had a Li detection, the rest being upper limits; 2 targets would be classified as Li-rich with $\mathrm{A(Li)_{non-LTE}}=1.54\pm0.21$ and $\mathrm{A(Li)_{non-LTE}}=1.60\pm0.21$. Their interpretation is that all of the giants except for the two Li-rich had experienced severe post-MS Li depletion, and that the two high measurements were stars with inhibited extra mixing. Our proposed explanation is very different: the two detections with high Li abundance are stars enriched in Li, possibly by planet engulfment (Section \ref{s_eng}), and the upper limits simply reflect FDU from a population that has experienced MS depletion. This does not preclude mixing on the RGB itself, since the sample consists of upper limits, but the data do not require it for the entire sample. To explain the differences in our interpretation, we begin by briefly summarizing their approach, which we then check in turn using our models.

S16 begin by comparing their abundances with those predicted by standard dredge-up theory, assuming a MS solar meteoritic abundance, and argue that the two Li-rich giants align with the predictions while all the other giants have Li well below them. They then argue that all the low luminosity stars in their sample are core-He burning stars, based on a combination of H-R diagram position relative to isochrones and CN measurements of the FDU. Within this framework, high Li is a sign of Li preservation, and all other stars must have experienced in situ mixing on the RGB, in addition to any MS mixing that could have occurred.

We instead argue that the data are better explained by a different interpretation of the evolutionary stages and mixing histories of the stars; namely, that both first ascent (inert He cores) and red clump (core He burning) stars are present and that we can reproduce the CN measurements without including thermohaline mixing. Instead of assuming a solar meteoritic zero point for Li, we use the average Li abundance found in the turn-off of other similar clusters. It would be natural to use the abundance of less evolved stars in Trumpler 20 as a zero-point, however anchoring models to the only such star reported (subgiant MG 430) is risky. Our models of MS depletion and standard FDU are broadly consistent with the Li upper limits of the Trumpler 20 giants as well as Li detections in other open cluster giants of similar mass. Within this framework, the two Li-rich giants therefore stand out as anomalously high, rather than normal, and we investigate planet engulfment as a possible origin.

\begin{figure*}[!htb]
\plottwo{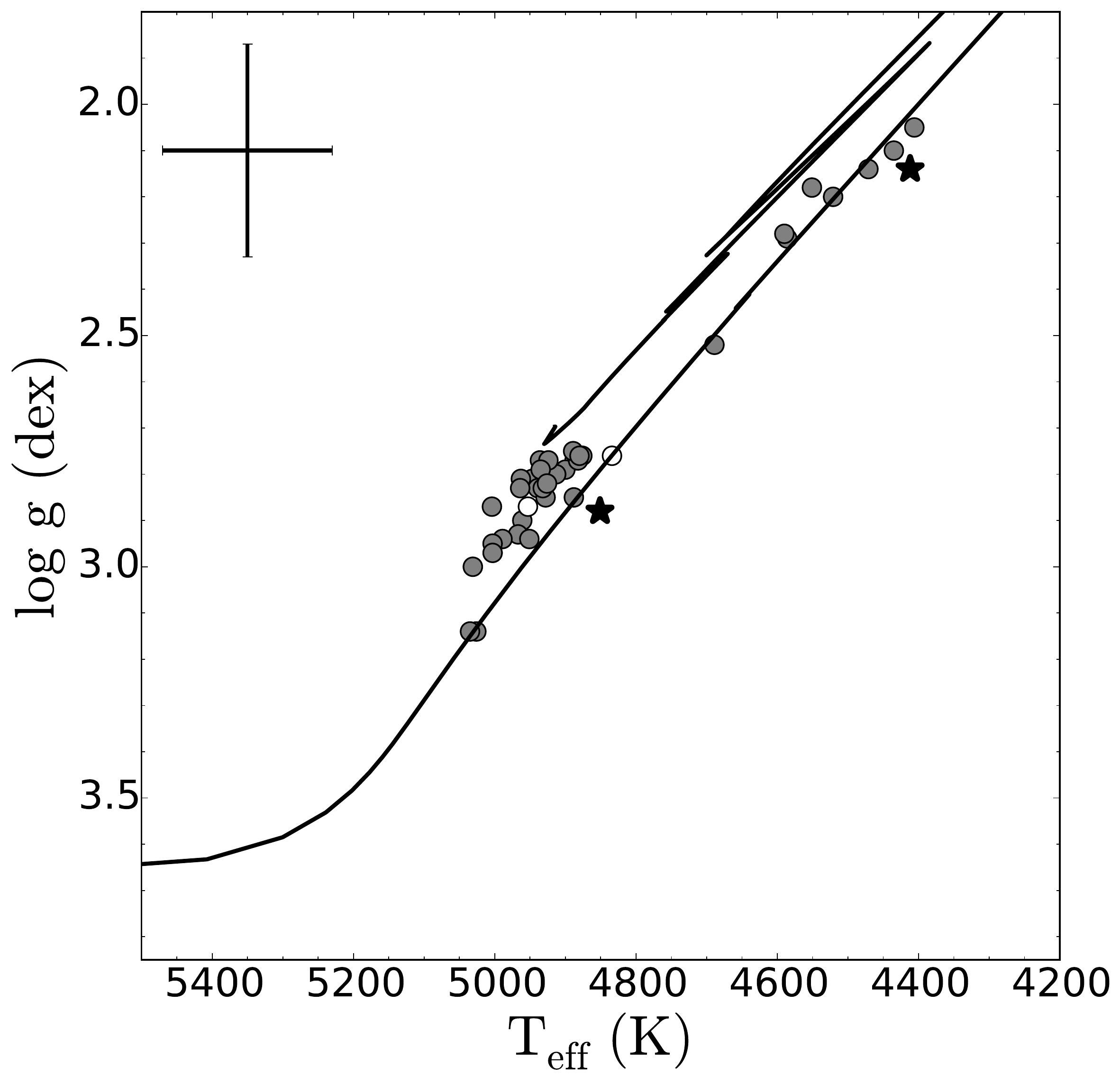}{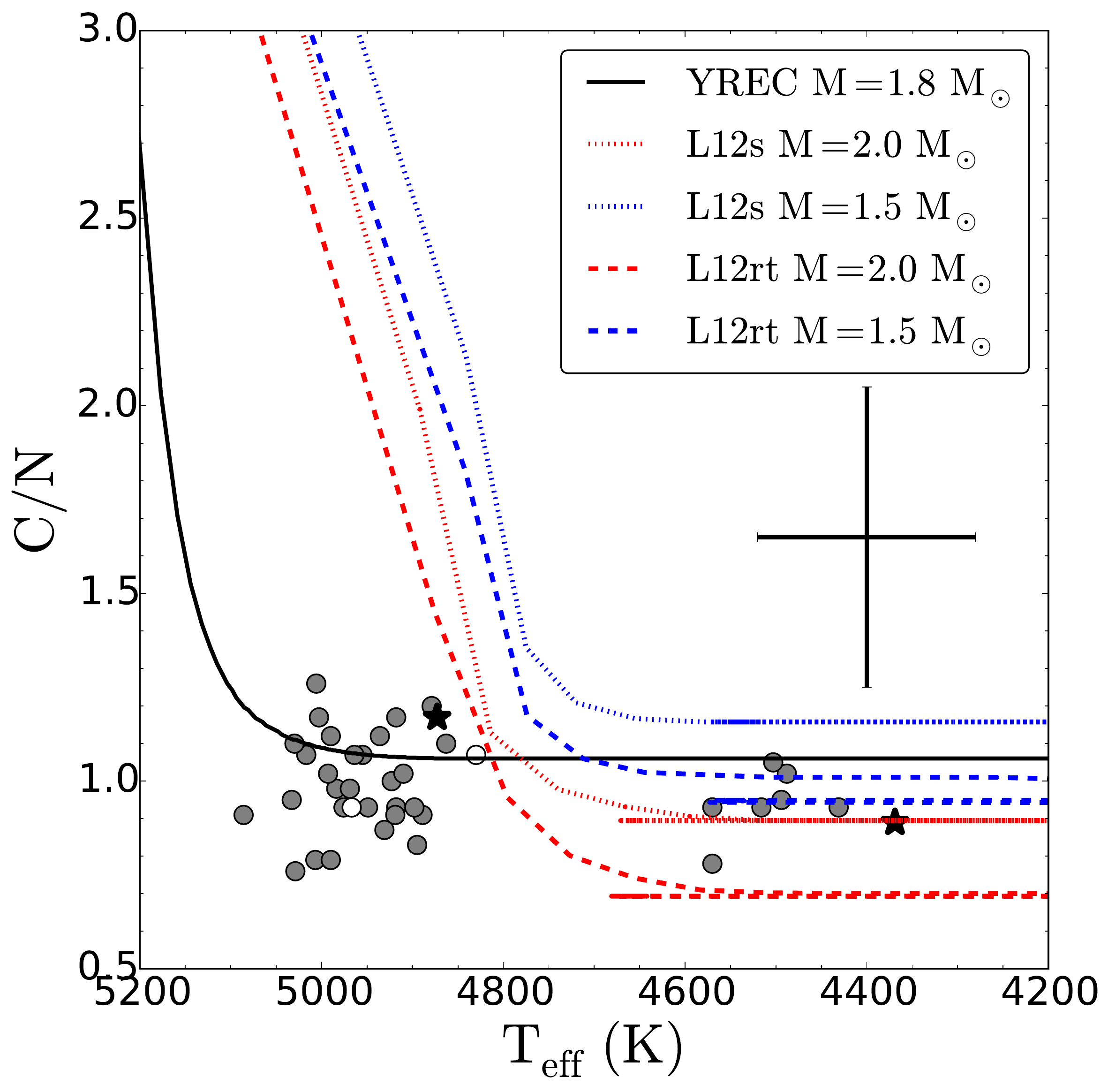}
\caption{Left panel: Trumpler 20 giants in the $\mathrm{T_{eff}}$-$\log g$ plane. Gray circles are giants with only Li upper limits, filled stars are the two Li-rich giants and open circles are stars with Li detections. These are plotted along a YREC isochrone with [Fe/H]$=+0.17$ and an age of $1.66$ Gyrs. Right panel: C/N abundance ratio as a function of $\mathrm{T_{eff}}$ for the giants in Trumpler 20, same symbols as left panel. The solid line shows the YREC model of a $1.8\,\mathrm{M_{\odot}}$ star with [Fe/H]=$+0.17$. For comparison, we show the $1.5\,\mathrm{M_{\odot}}$ (blue) and $2.0\,\mathrm{M_{\odot}}$ (red), solar metallicity standard models (L12s, dotted) and those with rotational and thermohaline mixing (L12rt, dashed) of \citet{lagarde12}, used by S16. The horizontal lines at lower $\mathrm{T_{eff}}$ in these models are due to the horizontal branch.\label{fig1}}
\end{figure*}

Our interpretation of the data is partially based on the evolutionary stage of the giants. Figure \ref{fig1}, left panel, shows the location of the stars in the $\mathrm{T_{eff}}$-$\mathrm{\log g}$ plane, and an $1.66$ Gyr isochrone with [Fe/H]$=+0.17$, obtained with the Yale Rotating Evolutionary Code \citep[YREC,][]{pinsonneault89,demarque08}, but the evolutionary stage is hard to identify from this information alone, as the group of giants with low $\log g$ could be either AGB or RGB stars, and the group with higher $\log g$ should include first ascent RGB and clump giants.

The C/N ratio can be useful to narrow down the evolutionary stage of the giants, as, during the FDU, carbon is reduced while nitrogen is increased, thus decreasing the C/N ratio. In Figure \ref{fig1}, right panel, all the giants have low C/N ratios, indicating that they have gone through FDU. The data is compared with models from \citet{lagarde12} with and without mixing along with predictions from standard evolutionary YREC models for a $1.8\,M_{\odot}$ star with [Fe/H]=$+0.17$ (black solid line). Those with mixing include rotational-induced mixing starting in the MS and thermohaline mixing after the RGB bump.

Note that the range of predicted C/N levels after FDU for all models (both with and without extra mixing) is about the same size as the typical uncertainty in the C/N measurements. Therefore, the C/N ratio is not a strong diagnostic of extra mixing in the Trumpler 20 giants.

We can see a shift in temperature between the YREC and \citet{lagarde12} models due to the use of a different mixing-length parameter. To fit the upper RGB of the cluster, our models suggest an $\alpha=2.1$, and considering the YREC model that predicts hotter temperatures for the base of the RGB, the giants are still consistent with first ascent RGB stars.

We can estimate the expected relative number of stars in both the RGB and core-He burning phases by relying on well-established theory. We compare the timescales that stars spend on those stages and share the same $\log g$ (from $\log g=3.2$ to $2.7$) in the high temperature group:
\begin{equation}
\frac{N_{RGB}}{N_{HB}}=\frac{t_{RGB}}{t_{HB}}.
\end{equation}
From the YREC models for the $1.8\,\mathrm{M_\odot}$ star with super-solar metallicity, $t_{RGB}\sim 7.5\times10^7$ yr, while for the core-He burning phase it is $t_{HB}\sim 1.1\times10^8$ yr. This means that $\sim19$ of the 31 giants in the high temperature group of this cluster should belong to the clump. This fraction is mainly illustrative since the exact timescales can change by using different assumptions. 
Nevertheless it challenges the assumption of S16 that all of the hotter stars must be red clump stars. In fact, the Li-rich star in this group stands out from the rest of the stars at same $\log g$ (Figure \ref{fig1}, left panel) further indicating that this could be an RGB star.

\begin{figure*}[!htb]
\begin{center}
\includegraphics[width=0.9\textwidth]{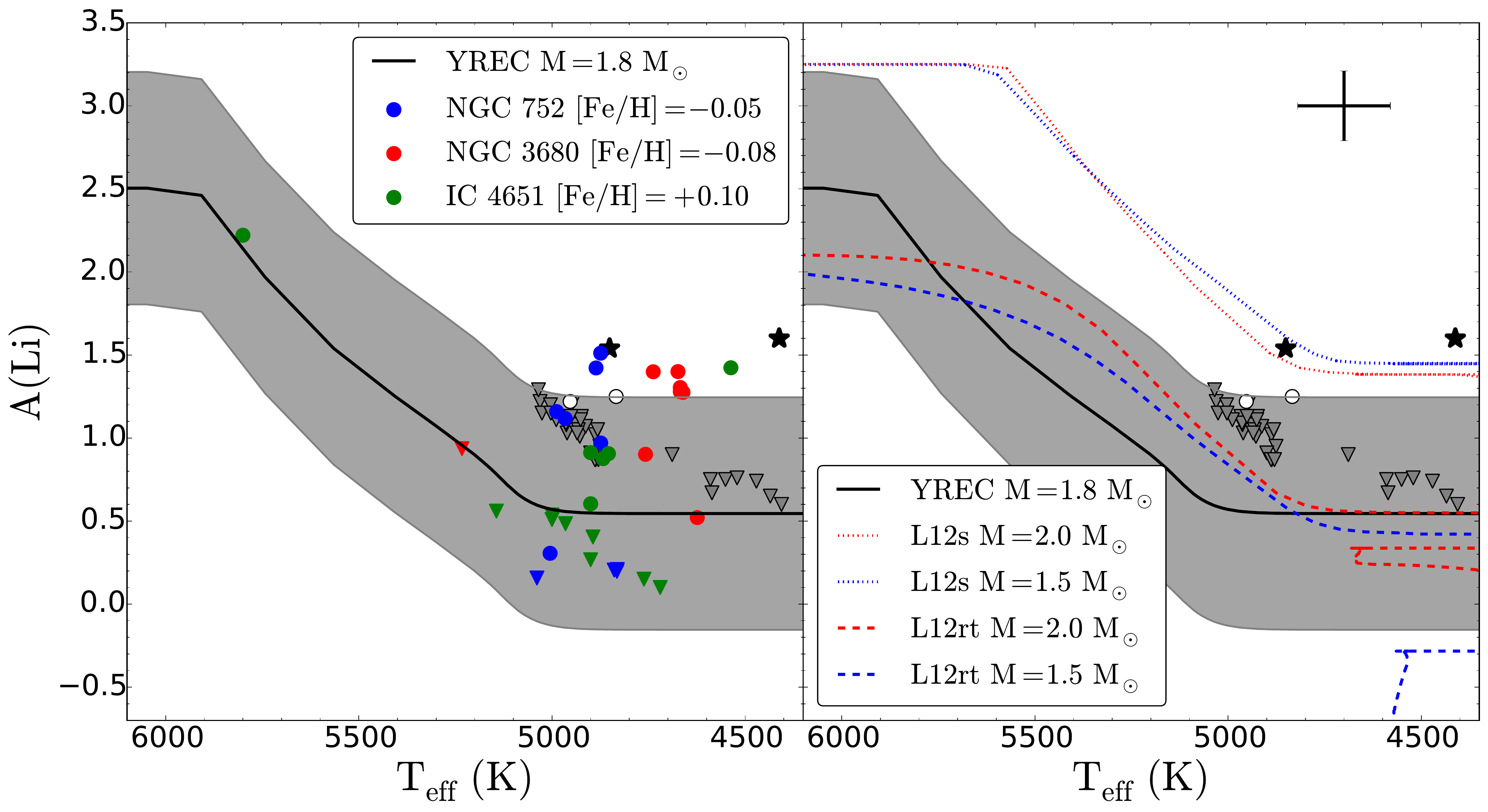}
\caption{$Gaia$-ESO Li abundances of Trumpler 20 giants as a function of $\mathrm{T_{eff}}$. Gray triangles are Li upper limits, open circles are detections and filled stars are Li-rich giants. The left panel shows the Trumpler 20 data and those of 3 other open clusters along with a YREC standard model, where the possible spread in MS abundance is shown as a gray band. In the right panel, the data is compared with the YREC and \citet{lagarde12} models (same styles and colors as Figure \ref{fig1}). The upper limits are consistent with YREC models once calibrated to the turn-off abundance of other similar open clusters. The non Li-rich giants are all consistent with an explanation of MS Li depletion and standard FDU with no extra-mixing required, which emphasizes the unusual nature of the Li-rich giants.}
\label{fig2}
\end{center}
\end{figure*}

The Li abundance of giants in Trumpler 20 can be seen in Figure \ref{fig2}, as a function of $\mathrm{T_{eff}}$. The left panel shows a comparison with our standard YREC model of a $1.8\,\mathrm{M_\odot}$ star and data from other open clusters of similar turn-off masses and different metallicities: NGC 752 \citep{bocektopcu15}, NGC 3680 \citep{pasquini01,DM16} and IC 4651 \citep{pasquini04, DM16}.
In the YREC model here, the MS Li abundance is calibrated to the turn-off abundance measured in these open clusters. Also, we use data from these other clusters to see how the Li in their giants compare to the upper limits in Trumpler 20.

IC 4651 has a very similar turn-off mass with super-solar metallicity [Fe/H]=$+0.10$, and its turn-off stars have a large spread in Li abundances from $\mathrm{A(Li)}\sim1.6$ to meteoritic values, possibly needing mixing to be explained \citep{pasquini04}. The same is observed in the open cluster NGC 3680 with similar age, but [Fe/H]$\sim-0.08$ \citep{AT09}, where Li abundances range from $\mathrm{A(Li)}\sim1.9$ to meteoritic \citep{pasquini01, AT09}. The similarly aged cluster NGC 752 with [Fe/H]$\sim-0.05$ \citep{AT09}, does not have enough data in the turn-off region to conclude anything. Overall the average Li abundance for turn-off stars is $\mathrm{A(Li)}=2.5\pm0.7$, which we used to calibrate the models and the possible internal spread found for a cluster (gray band in Figure \ref{fig2}).

Figure \ref{fig2} shows the Li abundance of the giants corrected by non-LTE using the grids from \citet{lind09li}. The Trumpler 20 giants are not special in any sense, and even some giants found in other clusters show similar Li abundance than the two Li-rich giants in Trumpler 20. These giants could be interesting to study given their high levels of Li when compared to other stars. On the other hand, the Li upper limits in the Trumpler 20 giants are similar to measurements in the comparison clusters, so we could think of them as marginal detections. It is still possible that some extra-mixing mechanism is being triggered past the RGB bump, as the Trumpler 20 giants that are located after that point show systematically lower Li abundances.

The fact that our estimation of Li abundance after FDU, including the corresponding zero-point uncertainty, agrees with the abundance of all giants in the sample except for the two Li-rich giants, argues in favor of this alternative interpretation of the observed Li pattern in the RGB of Trumpler 20.

\section{The two Li-rich giants in Trumpler 20: Possible engulfment}\label{s_eng}

We demonstrated in Section \ref{s_mixing} that all Trumpler 20 stars in the S16 sample (excepting the 2 Li-rich stars) are consistent with standard FDU after some MS Li depletion, supported by the $\mathrm{A(Li)}$ of turn-off stars in other similar clusters. Therefore, our interpretation of the Trumpler 20 stars suggests that the Li-rich giants have followed the same $\mathrm{A(Li)}$ evolution and were subsequently enriched in Li by a different mechanism, possibly the accretion of a substellar mass companion. A thorough analysis of the Li enrichment by this process was done in \citet{AG16}, where we model the engulfment, confirming that it can explain Li abundances up to $\mathrm{A(Li)}=2.2$ in giants, well above the level of the two Li-rich giants in Trumpler 20.

We model the engulfment of a substellar mass companion by the $1.8\,\mathrm{M_\odot}$ star, corresponding to the turn-off mass of Trumpler 20, with [Fe/H]=$+0.17$. The engulfment is produced at an age of $1.56$ Gyrs or $\mathrm{log\,g}\sim3.3$. Details on the parameters used in the YREC stellar evolutionary code or assumptions on the companion can be found in \citet{AG16}. Although MS extra-mixing was not considered there, by calibrating the Li abundance of the model in the MS we are taking into account processes that may act before the turn-off.

\begin{figure}[!htb]
\begin{center}
\includegraphics[width=0.4\textwidth]{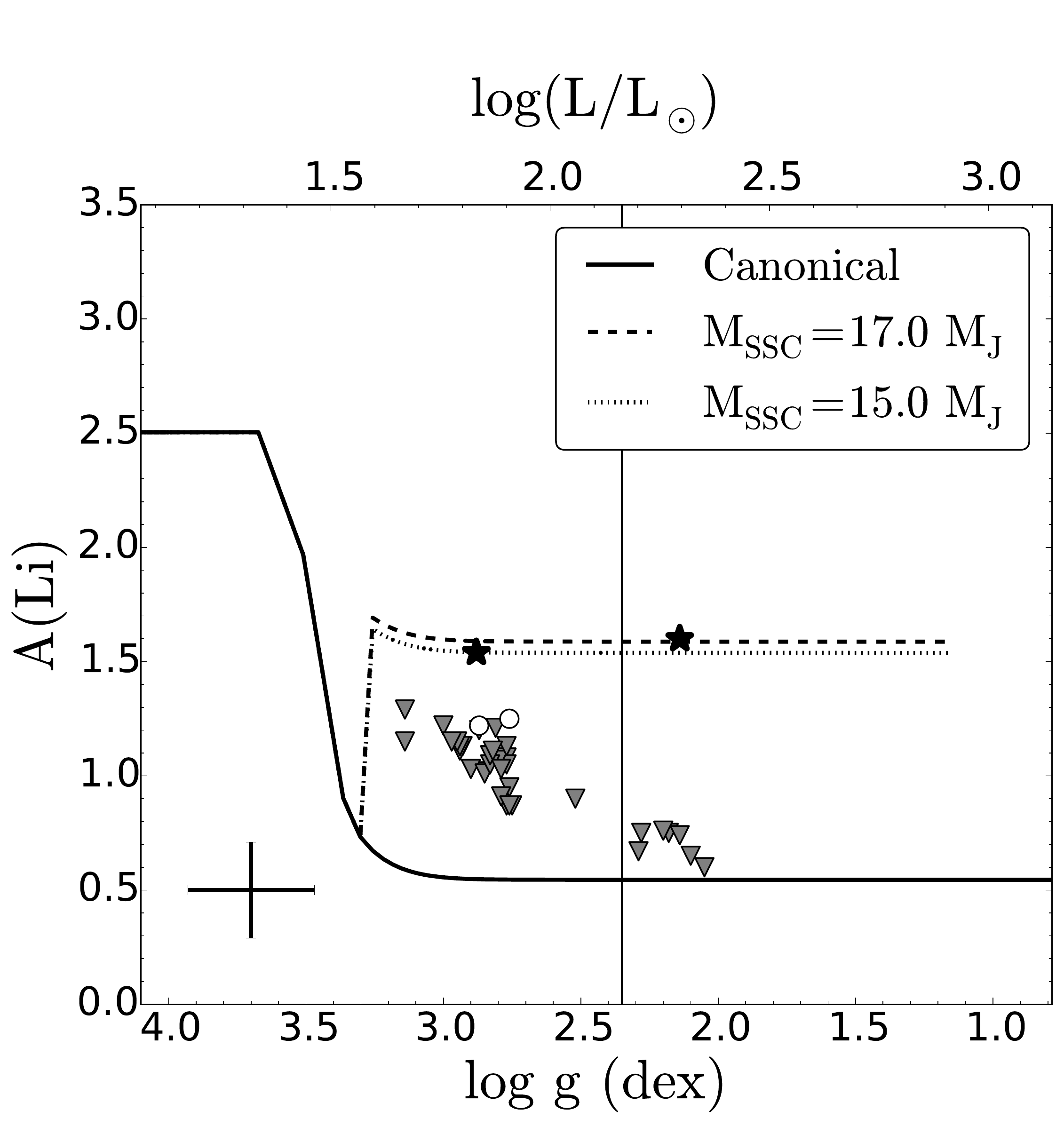}
\caption{Models of substellar companion engulfment to explain the Li-rich giants in Trumpler 20. The solid line is the same YREC standard model calibrated to the turn-off abundance in other similar clusters as shown in Figure \ref{fig2}. Dashed and dotted lines are models with engulfment of a $15\,\mathrm{M_J}$ and $17\,\mathrm{M_J}$ respectively at a $\mathrm{log\,g}\sim3.3$ corresponding to an age of the star of $1.56$ Gyrs. The vertical line shows the position of the RGB bump.}
\label{fig3}
\end{center}
\end{figure}

In Figure \ref{fig3} we show the evolution of surface $\mathrm{A(Li)}$ with and without engulfment. To explain the two Li-rich stars, the accretion of a substellar mass companion of $16^{+6}_{-10}\,\mathrm{M_J}$ is required. Changing the initial Li content of the star to consider the possible spread in turn-off abundances produces a negligible effect in the necessary companion mass compared to the uncertainty in Li measurements.

Figure \ref{fig3} shows two examples: $15\,\mathrm{M_J}$ (dashed line) and $17\,\mathrm{M_J}$ (dotted line). As one of the Li-rich giants is located after the RGB bump, indicated by the vertical line, an alternative formation scenario to account for its Li level is internal Li production plus enhanced extra-mixing. However, mixing mechanisms proposed so far that could increase Li seem to be less efficient at higher metallicities \citep{martell08}, as is the case of Trumpler 20.

Previously, \citet{siesslivio99} have studied the possibility of brown dwarf accretion, finding that not only should the surface Li abundance of the giant increase but also that the giant could spin up.
Accordingly, we note that although it is expected that the accretion of substellar mass companions by the giant increases its rotation rate, the low $vsini$ measured for the Li-rich stars is not enough to rule out this scenario. A rotational signal could last for a large portion of the RGB \citep{privitera16}, depending on several assumptions as the wind magnetic braking efficiency, so a more detailed analysis of this specific system would be needed.

To confirm our picture of Li evolution in Trumpler 20, it would be interesting to obtain Li abundance measurements of subgiants or turn-off stars in this cluster. This would inform us of the exact Li abundance and possible spread of stars before they go through the FDU. Then it would be possible to know if those giants at lower $\log g$ only had different turn-off Li abundances or if a process of extra mixing is triggered after the RGB bump, further decreasing the Li abundance.

Finally, the Li-rich giants in Trumpler 20 are compatible with a fraction of $1.8\,\mathrm{M_\odot}$ MS stars hosting close brown dwarf companions (that could be engulfed as the star approaches the RGB) of order $5\%$. This is intriguing, given suggestions that there is a higher fraction of massive planet hosts among A-dwarfs relative to solar analogs \citep[e.g.][]{johnson10}. However, observations show a lack of detected close-in planets around massive stars, making it necessary to study if this lack of objects is produced by engulfment when the star leaves the MS \citep[which is not supported by current models of tidal evolution,][]{kunitomo11}. Enhanced statistics for Li-rich giants in a larger sample, for example red giants with asteroseismic mass estimates, could therefore test the existence of the \textit{brown dwarf desert} \citep{gretherlineweaver06,troup16} in this mass domain. We stress that the current sample is small and that our results are suggestive rather than conclusive.

\section{Summary}

Trumpler 20 presents a great opportunity to study the problem of Li-rich giants in a sample of stars with the same age and very similar post turn-off masses. S16 have published atmospheric parameters and Li abundances for 40 giants, finding that 2 of them are Li-rich. Our interpretation of the same evidence is that most of the giants are consistent with MS depletion and standard dilution produced by FDU. We conclude that both Li-rich giants are unusual in nature and have been enriched in Li, possibly by the engulfment of substellar mass companions of $16^{+6}_{-10}\,\mathrm{M_J}$.
We cannot exclude the presence of mixing on the RGB, as most of the stars in the cluster only have Li upper limits, but this extra-mixing is not needed to explain the Li abundance pattern of the cluster.

Other authors have found samples where the Li-rich giants are thought to be produced by planet engulfment. \citet{casey16} present an interesting sample to study, although we are not certain that all of those giants are first ascent RGB and we think that the lack of close-in planets around subgiants is not a strong enough argument to point to planet engulfment, as it would predict a much larger amount of Li-rich giants before the bump.

The fraction of Li-rich giants in Trumpler 20, corresponding to $5\%$ is rather high compared to what is usually found (from $1\%$ to $2\%$). This fraction may indicate how many dwarfs of $1.8\,\mathrm{M_\odot}$ should have close-in substellar mass companions of high mass during the MS.

Further abundance measurements of these and other stars in this cluster, most preferably subgiants and turn-off stars, would allow a better study of the phenomenon producing the Li enrichment.

\acknowledgments
We thank the anonymous referee for a very useful report that helped improve the quality of the science presented in this work. We thank Luca Sbordone for helpful discussion, Jamie Tayar and Joel Zinn for guidance on horizontal branch models, and Elisa Delgado-Mena for comments to improve Figure 2 and its interpretation. Support for C.A.-G. is provided by CONICYT-PCHA Doctorado Nacional 2013-21130353. C.A.-G. and J.C. acknowledge support from the Chilean Ministry for the Economy, Development, and Tourism's Programa Iniciativa Cient\'ifica Milenio, through grant IC120009 awarded to the Millenium Institute of Astrophysics (MAS) and from PFB-06 Centro de Astronomia y Tecnologias Afines. M.H.P. acknowledges support from NASA grant NNX15AF13G. J.K.C. was supported by an appointment to the NASA Postdoctoral Program at the Goddard Space Flight Center, administered by Universities Space Research Association under contract with NASA.

\bibliography{Tr20_engulfment}{}

\begin{thebibliography}{}
\expandafter\ifx\csname natexlab\endcsname\relax\def\natexlab#1{#1}\fi

\bibitem[{{Aguilera-G{\'o}mez} {et~al.}(2016){Aguilera-G{\'o}mez},
  {Chanam{\'e}}, {Pinsonneault}, \& {Carlberg}}]{AG16}
{Aguilera-G{\'o}mez}, C., {Chanam{\'e}}, J., {Pinsonneault}, M.~H., \&
  {Carlberg}, J.~K. 2016, \apj, 829, 127

\bibitem[{{Alexander}(1967)}]{alexander67}
{Alexander}, J.~B. 1967, The Observatory, 87, 238

\bibitem[{{Anthony-Twarog} {et~al.}(2009){Anthony-Twarog}, {Deliyannis},
  {Twarog}, {Croxall}, \& {Cummings}}]{AT09}
{Anthony-Twarog}, B.~J., {Deliyannis}, C.~P., {Twarog}, B.~A., {Croxall},
  K.~V., \& {Cummings}, J.~D. 2009, \aj, 138, 1171

\bibitem[{{B{\"o}cek Topcu} {et~al.}(2015){B{\"o}cek Topcu}, {Af{\c s}ar},
  {Schaeuble}, \& {Sneden}}]{bocektopcu15}
{B{\"o}cek Topcu}, G., {Af{\c s}ar}, M., {Schaeuble}, M., \& {Sneden}, C. 2015,
  \mnras, 446, 3562

\bibitem[{{Brown} {et~al.}(1989){Brown}, {Sneden}, {Lambert}, \&
  {Dutchover}}]{brown89}
{Brown}, J.~A., {Sneden}, C., {Lambert}, D.~L., \& {Dutchover}, Jr., E. 1989,
  \apjs, 71, 293

\bibitem[{{Cameron} \& {Fowler}(1971)}]{cameronfowler71}
{Cameron}, A.~G.~W., \& {Fowler}, W.~A. 1971, \apj, 164, 111

\bibitem[{{Casey} {et~al.}(2016){Casey}, {Ruchti}, {Masseron}, {Randich},
  {Gilmore}, {Lind}, {Kennedy}, {Koposov}, {Hourihane}, {Franciosini}, {Lewis},
  {Magrini}, {Morbidelli}, {Sacco}, {Worley}, {Feltzing}, {Jeffries},
  {Vallenari}, {Bensby}, {Bragaglia}, {Flaccomio}, {Francois}, {Korn},
  {Lanzafame}, {Pancino}, {Recio-Blanco}, {Smiljanic}, {Carraro}, {Costado},
  {Damiani}, {Donati}, {Frasca}, {Jofr{\'e}}, {Lardo}, {de Laverny}, {Monaco},
  {Prisinzano}, {Sbordone}, {Sousa}, {Tautvai{\v s}ien{\.e}}, {Zaggia},
  {Zwitter}, {Delgado Mena}, {Chorniy}, {Martell}, {Silva Aguirre}, {Miglio},
  {Chiappini}, {Montalban}, {Morel}, \& {Valentini}}]{casey16}
{Casey}, A.~R., {Ruchti}, G., {Masseron}, T., {et~al.} 2016, \mnras, 461, 3336

\bibitem[{{Delgado Mena} {et~al.}(2016){Delgado Mena}, {Tsantaki}, {Sousa},
  {Kunitomo}, {Adibekyan}, {Zaworska}, {Santos}, {Israelian}, \&
  {Lovis}}]{DM16}
{Delgado Mena}, E., {Tsantaki}, M., {Sousa}, S.~G., {et~al.} 2016, \aap, 587,
  A66

\bibitem[{{Demarque} {et~al.}(2008){Demarque}, {Guenther}, {Li}, {Mazumdar}, \&
  {Straka}}]{demarque08}
{Demarque}, P., {Guenther}, D.~B., {Li}, L.~H., {Mazumdar}, A., \& {Straka},
  C.~W. 2008, \apss, 316, 31

\bibitem[{{Donati} {et~al.}(2014){Donati}, {Cantat Gaudin}, {Bragaglia},
  {Friel}, {Magrini}, {Smiljanic}, {Vallenari}, {Tosi}, {Sordo}, {Tautvai{\v
  s}ien{\.e}}, {Blanco-Cuaresma}, {Costado}, {Geisler}, {Klutsch}, {Mowlavi},
  {Mu{\~n}oz}, {San Roman}, {Zaggia}, {Gilmore}, {Randich}, {Bensby},
  {Flaccomio}, {Koposov}, {Korn}, {Pancino}, {Recio-Blanco}, {Franciosini}, {de
  Laverny}, {Lewis}, {Morbidelli}, {Prisinzano}, {Sacco}, {Worley},
  {Hourihane}, {Jofr{\'e}}, {Lardo}, \& {Maiorca}}]{donati14}
{Donati}, P., {Cantat Gaudin}, T., {Bragaglia}, A., {et~al.} 2014, \aap, 561,
  A94

\bibitem[{{Gratton} {et~al.}(2000){Gratton}, {Sneden}, {Carretta}, \&
  {Bragaglia}}]{gratton00}
{Gratton}, R.~G., {Sneden}, C., {Carretta}, E., \& {Bragaglia}, A. 2000, \aap,
  354, 169

\bibitem[{{Grether} \& {Lineweaver}(2006)}]{gretherlineweaver06}
{Grether}, D., \& {Lineweaver}, C.~H. 2006, \apj, 640, 1051

\bibitem[{{Johnson} {et~al.}(2010){Johnson}, {Aller}, {Howard}, \&
  {Crepp}}]{johnson10}
{Johnson}, J.~A., {Aller}, K.~M., {Howard}, A.~W., \& {Crepp}, J.~R. 2010,
  \pasp, 122, 905

\bibitem[{{Kumar} {et~al.}(2011){Kumar}, {Reddy}, \& {Lambert}}]{kumar11}
{Kumar}, Y.~B., {Reddy}, B.~E., \& {Lambert}, D.~L. 2011, \apjl, 730, L12

\bibitem[{{Kunitomo} {et~al.}(2011){Kunitomo}, {Ikoma}, {Sato}, {Katsuta}, \&
  {Ida}}]{kunitomo11}
{Kunitomo}, M., {Ikoma}, M., {Sato}, B., {Katsuta}, Y., \& {Ida}, S. 2011,
  \apj, 737, 66

\bibitem[{{Lagarde} {et~al.}(2012){Lagarde}, {Decressin}, {Charbonnel},
  {Eggenberger}, {Ekstr{\"o}m}, \& {Palacios}}]{lagarde12}
{Lagarde}, N., {Decressin}, T., {Charbonnel}, C., {et~al.} 2012, \aap, 543,
  A108

\bibitem[{{Lebzelter} {et~al.}(2012){Lebzelter}, {Uttenthaler}, {Busso},
  {Schultheis}, \& {Aringer}}]{lebzelter12}
{Lebzelter}, T., {Uttenthaler}, S., {Busso}, M., {Schultheis}, M., \&
  {Aringer}, B. 2012, \aap, 538, A36

\bibitem[{{Lind} {et~al.}(2009{\natexlab{a}}){Lind}, {Asplund}, \&
  {Barklem}}]{lind09li}
{Lind}, K., {Asplund}, M., \& {Barklem}, P.~S. 2009{\natexlab{a}}, \aap, 503,
  541

\bibitem[{{Lind} {et~al.}(2009{\natexlab{b}}){Lind}, {Primas}, {Charbonnel},
  {Grundahl}, \& {Asplund}}]{lind09}
{Lind}, K., {Primas}, F., {Charbonnel}, C., {Grundahl}, F., \& {Asplund}, M.
  2009{\natexlab{b}}, \aap, 503, 545

\bibitem[{{Martell} {et~al.}(2008){Martell}, {Smith}, \& {Briley}}]{martell08}
{Martell}, S.~L., {Smith}, G.~H., \& {Briley}, M.~M. 2008, \aj, 136, 2522

\bibitem[{{Monaco} {et~al.}(2014){Monaco}, {Boffin}, {Bonifacio}, {Villanova},
  {Carraro}, {Caffau}, {Steffen}, {Ahumada}, {Beletsky}, \&
  {Beccari}}]{monaco14}
{Monaco}, L., {Boffin}, H.~M.~J., {Bonifacio}, P., {et~al.} 2014, \aap, 564, L6

\bibitem[{{Pasquini} {et~al.}(2001){Pasquini}, {Randich}, \&
  {Pallavicini}}]{pasquini01}
{Pasquini}, L., {Randich}, S., \& {Pallavicini}, R. 2001, \aap, 374, 1017

\bibitem[{{Pasquini} {et~al.}(2004){Pasquini}, {Randich}, {Zoccali}, {Hill},
  {Charbonnel}, \& {Nordstr{\"o}m}}]{pasquini04}
{Pasquini}, L., {Randich}, S., {Zoccali}, M., {et~al.} 2004, \aap, 424, 951

\bibitem[{{Pinsonneault} {et~al.}(1989){Pinsonneault}, {Kawaler}, {Sofia}, \&
  {Demarque}}]{pinsonneault89}
{Pinsonneault}, M.~H., {Kawaler}, S.~D., {Sofia}, S., \& {Demarque}, P. 1989,
  \apj, 338, 424

\bibitem[{{Privitera} {et~al.}(2016){Privitera}, {Meynet}, {Eggenberger},
  {Vidotto}, {Villaver}, \& {Bianda}}]{privitera16}
{Privitera}, G., {Meynet}, G., {Eggenberger}, P., {et~al.} 2016, ArXiv
  e-prints, arXiv:1606.08027

\bibitem[{{Randich} {et~al.}(2013){Randich}, {Gilmore}, \& {Gaia-ESO
  Consortium}}]{randichgilmore13}
{Randich}, S., {Gilmore}, G., \& {Gaia-ESO Consortium}. 2013, The Messenger,
  154, 47

\bibitem[{{Reddy} \& {Lambert}(2016)}]{reddylambert16}
{Reddy}, A.~B.~S., \& {Lambert}, D.~L. 2016, \aap, 589, A57

\bibitem[{{Sackmann} \& {Boothroyd}(1992)}]{sackmannboothroyd92}
{Sackmann}, I.-J., \& {Boothroyd}, A.~I. 1992, \apjl, 392, L71

\bibitem[{{Siess} \& {Livio}(1999)}]{siesslivio99}
{Siess}, L., \& {Livio}, M. 1999, \mnras, 308, 1133

\bibitem[{{Silva Aguirre} {et~al.}(2014){Silva Aguirre}, {Ruchti}, {Hekker},
  {Cassisi}, {Christensen-Dalsgaard}, {Datta}, {Jendreieck}, {Jessen-Hansen},
  {Mazumdar}, {Mosser}, {Stello}, {Beck}, \& {de Ridder}}]{SA14}
{Silva Aguirre}, V., {Ruchti}, G.~R., {Hekker}, S., {et~al.} 2014, \apjl, 784,
  L16

\bibitem[{{Smiljanic} {et~al.}(2016){Smiljanic}, {Franciosini}, {Randich},
  {Magrini}, {Bragaglia}, {Pasquini}, {Vallenari}, {Tautvai{\v s}ien{\.e}},
  {Biazzo}, {Frasca}, {Donati}, {Delgado Mena}, {Casey}, {Geisler},
  {Villanova}, {Tang}, {Sousa}, {Gilmore}, {Bensby}, {Fran{\c c}ois},
  {Koposov}, {Lanzafame}, {Pancino}, {Recio-Blanco}, {Costado}, {Hourihane},
  {Lardo}, {de Laverny}, {Lewis}, {Monaco}, {Morbidelli}, {Sacco}, {Worley},
  {Zaggia}, \& {Martell}}]{smiljanic16}
{Smiljanic}, R., {Franciosini}, E., {Randich}, S., {et~al.} 2016, \aap, 591,
  A62

\bibitem[{{Tautvai{\v s}ien{\.e}} {et~al.}(2015){Tautvai{\v s}ien{\.e}},
  {Drazdauskas}, {Mikolaitis}, {Barisevi{\v c}ius}, {Puzeras}, {Stonkut{\.e}},
  {Chorniy}, {Magrini}, {Romano}, {Smiljanic}, {Bragaglia}, {Carraro}, {Friel},
  {Morel}, {Pancino}, {Donati}, {Jim{\'e}nez-Esteban}, {Gilmore}, {Randich},
  {Jeffries}, {Vallenari}, {Bensby}, {Flaccomio}, {Recio-Blanco}, {Costado},
  {Hill}, {Jofr{\'e}}, {Lardo}, {de Laverny}, {Masseron}, {Moribelli}, {Sousa},
  \& {Zaggia}}]{tautvaisiene15}
{Tautvai{\v s}ien{\.e}}, G., {Drazdauskas}, A., {Mikolaitis}, {\v S}., {et~al.}
  2015, \aap, 573, A55

\bibitem[{{Troup} {et~al.}(2016){Troup}, {Nidever}, {De Lee}, {Carlberg},
  {Majewski}, {Fernandez}, {Covey}, {Chojnowski}, {Pepper}, {Nguyen},
  {Stassun}, {Nguyen}, {Wisniewski}, {Fleming}, {Bizyaev}, {Frinchaboy},
  {Garc{\'{\i}}a-Hern{\'a}ndez}, {Ge}, {Hearty}, {Meszaros}, {Pan}, {Allende
  Prieto}, {Schneider}, {Shetrone}, {Skrutskie}, {Wilson}, \&
  {Zamora}}]{troup16}
{Troup}, N.~W., {Nidever}, D.~L., {De Lee}, N., {et~al.} 2016, \aj, 151, 85

\bibitem[{{Wallerstein} \& {Sneden}(1982)}]{wallersteinsneden82}
{Wallerstein}, G., \& {Sneden}, C. 1982, \apj, 255, 577

\end{thebibliography}
\bibliographystyle{aasjournal}

\end{document}